# LUMINOUS LONG PERIOD VARIABLES IN GLOBULAR CLUSTERS AND THE GALACTIC BULGE: THEIR DEPENDENCE ON METALLICITY

*Jay A. Frogel*[1]

Department of Astronomy, The Ohio State University
174 West 18th Avenue, Columbus, Ohio  43210
e-mail:  frogel@galileo.mps.ohio-state.edu

and

*Patricia A. Whitelock*

South African Astronomical Observatory
P.O. Box 9, 7935 Observatory, Republic of South Africa
e-mail:  paw@saao.ac.za

---

[1] Also Visiting Senior Research Fellow, Department of Physics, University of Durham, England.

## ABSTRACT

We derive the frequency of occurrence of luminous long period variables (LLPVs) in globular clusters and in the Baade's Window field of the Galactic bulge. LLPVs occur only in clusters with [Fe/H] $\gtrsim -1.0$. In these clusters their frequency of occurrence relative to the number of giant stars appears to be independent of metallicity. Integrated over all metallicities, Baade's Window appears to be deficient in LLPVs. We estimate [Fe/H] values for Baade's Window LLPVs from their period and a log P vs. [Fe/H] relation, derived from the cluster variables, and find that this deficiency is due primarily to the absence of LLPVs with [Fe/H] $\gtrsim$ 0.0. We propose that this is due to enhanced mass loss rates in these LLPVs with a consequently abbreviated lifetime compared to lower metallicity LLPVs. A typical lifetime for cluster LLPVs is about $3 \times 10^5$ yrs. Finally, we emphasize the need for a much more complete survey for LLPVs in globular clusters.

Subject headings: stars: long-period variables — stars: evolution — stars: AGB — globular clusters: general — galaxies: stellar content — galaxies: the Galaxy

# 1. INTRODUCTION

Long Period Variables (LPVs), as defined by Payne-Gaposhkin & Gaposhkin (1938), Merrill (1940), and Feast (1973) are large amplitude, red variables usually of spectral type Me (but can also be of S or N type) with periods greater than 50 days but more typically of several hundred days and light amplitudes of several magnitudes in the visible. LPVs are all most likely on the asymptotic giant branch (AGB) and thus represent the final stage of giant evolution for low to intermediate mass stars with [Fe/H] $\gtrsim$ −1.0. They are found in a wide variety of environments including globular clusters and the Galactic bulge. Their frequency of occurrence in a stellar population is of interest for understanding the variables themselves and the late stage of evolution they represent, for estimating the rate of return of processed material to the interstellar medium, and, potentially, as indicators of the age and metallicity of the stellar environment in which they are found. In this paper we analyze existing observations of LPVs in globular clusters and the Galactic bulge with the specific goals of determining their frequency of occurrence and ages in two quite different populations as a function of metallicity.

It is difficult to determine with any degree of accuracy physical parameters for LPVs in the general field because of uncertainties in their distance moduli, reddening, and metallicity. For globular clusters, on the other hand, these parameters are relatively well known. Frogel and Elias (1988) demonstrated that *all* variables in globular clusters with $M_{bol}$ brighter than the predicted level of core He flash are usually classified as LPVs or, in H. S. Hogg's (1973) notation, M type. Since the periods of these stars almost all seem to be greater than 190 days and they have large amplitude light variations, they can further be classified as Mira variables. A search of Hogg's catalogue shows that such variables are found *only* in relatively metal rich globulars, not metal poor ones. Furthermore, the earlier work of Frogel, Persson, & Cohen (1983) based on photometry of luminous giants in more than 30 globular clusters and scans of a subset of these clusters in the near-IR, demonstrated that no other stars, either variable or non-variable, are more luminous than the level of core He flash. Thus, to paraphrase one of Frogel & Elias' main conclusions, all stars brighter than core He flash in Galactic globular clusters (i.e. the most luminous stars in these clusters) are variables; they are found only in relatively metal-rich clusters; and, finally, the characteristics of the variability indicates that they are all LPVs. For specificity we will call these stars luminous LPVs (LLPVs) to clearly distinguish them from semi-regular variables (SR type) which on occasion are referred to as LPVs as well. Nevertheless, for all cases known so far, there is a one-to-one correspondence between the Me or LPV classification in Hogg (1973) and the infrared characteristics delineated by Frogel & Elias. In the rest of this paper we will assume that this correspondence is always true.

Three near-IR photometric characteristics of LLPVs make it easy to distinguish them from all other stars in a globular cluster (Frogel 1985a; also see Feast *et al.* 1982, Menzies & Whitelock 1985). These characteristics are: 1) Their mean bolometric luminosities are always greater than the termination point (location of core He flash) of the first giant branch; as noted above, no other globular cluster stars are as luminous. 2) Their *JHK* colors are distinct from those of all other cluster giants due primarily to the effect of $H_2O$ absorption on the flux in the band-pass of the *H* filter. 3) At the same *J*–*K* color an LLPV has a markedly redder *K*–*L* color than a non-LLPV; this could arise from excess thermal circumstellar emission in the *L* band-pass (*e.g.* Frogel & Elias 1988). In addition to these three near-IR characteristics, the optical spectra of many LPVs including the LLPVs have strong emission lines at certain phases of their light cycles (*e.g.* Merrill 1940, Feast 1963). These emission lines result from shock waves produced by the pulsation.

In section 2 of this paper we derive measures for the frequency of occurrence of LLPVs in globular clusters. One is the ratio of bright ($M_{bol} \leq -1.2$) giants to LLPVs, while another is the ratio of the number of LLPVs to the integrated absolute $K$ magnitude of the cluster. These parameters are then used in section 3 to examine the relationship between bright giants and LLPVs in the Galactic bulge. Section 4 summarizes the results and presents an estimate for LLPV lifetimes.

## 2. LUMINOUS LPVS IN GLOBULAR CLUSTERS

In this section we derive estimates for the number of LLPVs in a cluster relative to both the number of bright giants and the integrated luminosity of bright giants. These ratios will then be used to estimate the lifetimes of LLPVs and to investigate the dependence of their frequency of occurrence on cluster metallicity. We will use our assumption of a one-to-one correspondence between near-IR properties that define an LLPV and optical LPV classification.

### 2.1 How to Count Luminous LPVs in Globular Clusters

NGC 6712 with an [Fe/H] = −1.01 on the scale of Zinn & West (1984) is the most metal poor cluster known to have an LLPV. In all, there are only 12 globular clusters with metallicities comparable to or greater than that of NGC 6712 which have been studied thoroughly enough, particularly in their cores, that any LLPVs in them should have been found.[2] These 12 clusters are listed in Table 1. Nine of them have a total of 19 LLPVs while three have none. In the past decade the only addition to the list of clusters with [Fe/H] ≥ −1 which do not appear to have any LLPVs is NGC6362 (Clement, Dickens, & Bingham 1995). Infrared photometry is available for all 19 LLPVs (Frogel, Cohen, & Persson 1983; Menzies & Whitelock 1985; Frogel 1985b; Frogel & Elias 1988).

Differences between Harris' (1987) compilation of numbers of variables of different types in clusters and the entries in Table 1 arise because there are a few cases where a star *tentatively* classified as an LPV from optical data did not satisfy the three defining criteria listed in the Introduction for an LLPV (thus we would conclude that the optical classification is in error), or where stars with limited optical data were, on the basis of near-IR data, clearly shown to be an LLPV. SRd type variables were generally excluded from the LLPV category as their bolometric magnitudes place them below the tip of the RGB. Feast (1981) and Lloyd Evans (1983) have suggested that such stars may be the metal poor ([Fe/H] ≤ −1.0) analogues of the LPVs found in metal rich globular clusters. Finally, the lack of adequate infrared and/or optical data for some of the clusters has also resulted in differences between the present compilation and that of Harris.

### 2.2 How to Count Non-Variable Giants in Globular Clusters

We now want an estimate of the relative frequency of LLPVs in globular clusters. The ratio of number of LLPVs to non-variable giants in a cluster can provide a useful measure of this relative frequency. While for the clusters in Table 1 we can be reasonably certain that all of the LLPVs have been counted, one cannot simply refer to the color-magnitude diagrams of the clusters in Table 1 and count non-variable giants. One reason is that for the more populous clusters such as NGC 104 (47 Tucanae) and NGC 5927, photometry of non-variable giants has been done only in selected parts of the clusters; their central regions are usually excluded from photometric studies. Also, for the sparser clusters with stellar photometry that does go into their centers, there is still substantial incompleteness, even for the brightest stars, probably due to a combination of crowding and saturation.

---

[2] Although many near-IR color-magnitude diagrams of globular clusters are now being published from which it should be possible to identify LLPV candidates with only a single epoch observation, observations of the crowded central regions of the clusters are still incomplete.

As an alternative approach we used a cluster's integrated $K$ magnitude as an indicator of its *relative* numbers of giants. This is a satisfactory approach since at least 50% of an old stellar system's $K$ light comes from giant stars (*e.g.* Fig. 7 of Frogel (1988) and Fig. 1.5 of Renzini (1994)). Integrated $V$ magnitudes, reddening values, and distance moduli were taken from the compilation of Harris (1996) except for NGC6712, values for which are from Kuchinski & Frogel (1995). These $V$ magnitudes were converted to absolute $K$ magnitudes with the integrated $V-K$ colors and magnitudes from unpublished data by M. Aaronson for all but 4 of the clusters. For these 4 - NGC 104, 5927, 6352, and 6362 – we used the following to get $(V-K)_0$ :

$$(V - K)_0 = 1.34 + 1.22 \times (U - V)_0 \qquad\qquad 1.$$

This is based on Aaronson *et al.* (1978) and unpublished data. (*U—V*) colors are from Harris (1996).

With the integrated absolute $K$ magnitudes for the 12 clusters in Table 1 we can proceed to estimate the LLPV frequency relative to the giant star population as a whole. Consider the following measure of the specific frequency of occurrence of LLPVs:

$$10^4 \sum N(LLPV) \Big/ \sum \left( 10^{-0.4 \times M_{K_0}} \right) \qquad\qquad 2.$$

where $N(LLPV)$ is the number of LLPVs in each cluster (col. 6 of Table 1), $M_K$ is the integrated absolute $K$ magnitude of a cluster, and the summations are over all clusters in Table 1. The value of this quantity is 0.74. The value of the denominator ($\times 10^{-3}$) for each cluster is given in the last column of Table 1 and, as mentioned in the previous paragraph, should be a good indicator of the *relative* number of giants in the clusters because more than half of the K light is from giant stars.

Next we want to be somewhat more precise than saying that 50% of a cluster's $K$ light is contributed by giants. A simple way to do this is with the stellar synthesis models of Aaronson *et al.* (1978) as modified by Frogel, Persson, & Cohen (1980). These models start with the Ciardullo & Demarque (1977) single burst isochrones and a 30% He abundance by mass. Four modifications were made to the models by Frogel *et al.* (1980): The locations of the isochrones in the L, $T_{eff}$ plane are sensitive to the scale height to mixing length ratio. Observations of individual globular cluster stars in the near-IR indicated that these locations needed to be shifted by a small amount. The color - effective temperature scale was modified to agree better with lunar occultation measurements of cool giants. Third, the metallicity scale was adjusted to agree with the best data available to Frogel *et al.* (1980) which, in fact, is quite close to the globular metallicity scale generally used today. Finally a numerical error in the Ciardullo & Demarque (1977) bolometric luminosity function was corrected.

Although the Yale tracks have been redone with new and improved input physics since the Ciardullo & Demarque (1977) isochrones were published 20 years ago, the models based on them that Frogel *et al.* (1980) used closely reproduce the dependence of the integrated *UBVJHK* colors of globular clusters on metallicity and also are consistent with studies of individual giants in the clusters (Frogel *et al.* 1983). A physically plausible range in the mass loss rate has only a small effect on the model prediction and none at all on the use to which they are put in this paper, especially given the small number statistics we are dealing with (19 LLPVs in 12 clusters). Thus for the purposes of this paper these models are adequate.

For purposes of illustration we take an old system with 1000 stars and Z=0.01. Details of this model are in Table 2. Column 1 lists the s values where s=2.35 is the Salpeter value for the slope of the main sequence mass function. Cols. 2 and 3 are the integrated $V-K$ color and absolute K magnitudes of the models. Cols. 4 and 5 show the percent of the total $K$ light that comes from stars with $M_{bol} \leq -1.2$ (already calculated by Frogel & Whitford 1987[3]) and the integrated $K$ magnitude of these bright giants alone. The sixth column gives the predicted numbers of LLPVs based on their frequency of occurrence (0.74) relative to the total $K$ light of the model (col. 3). Col. 7 gives the actual number of giants in the models with $M_{bol} \leq -1.2$ while col. 8 shows the percentage of giants that are on the AGB. Finally, col. 9 gives the number of giants with $M_{bol} \leq -1.2$ per LLPV. For comparison, if we chose the Z=0.001 model, 51 and 39% of the K light would be contributed by giants with $M_{bol} \leq -1.2$ for s = 0.0 and 2.35, respectively, while for the rather uncertain z=0.04 model the percentages are 59 and 51, respectively. Thus the fraction of $K$ light from bright giants is not particularly sensitive to the details of the model chosen for the range of [Fe/H] of interest. This being the case, we average the properties of the two models in Table 2. Thus, *our semi-theoretical estimate for the number of giants stars with $M_{bol} \leq -1.2$ per LLPV in metal rich globular clusters is 62.* This number is not very sensitive to the details of the model chosen. Furthermore, since an average of 48% of the $K$ light is coming from these bright giants, we will use 1.55 (=0.74/0.48) as the frequency of occurrence of LLPVs relative to the integrated $K$ luminosity of these giants in the sense defined above.

## 3. APPLICATION TO THE GALACTIC BULGE

For metal rich globular clusters we have estimated the ratio of the number of GB and AGB stars (with $M_{bol} \leq -1.2$) to LLPVs to be 62 to 1. If stars in the Galactic bulge are of similar age to the globular clusters in Table 1, one might expect the frequency of occurrence of LLPVs in the two populations to be similar, that is if no differences arose due to differences in the [Fe/H] distribution of the two populations. We will use the well-studied stars in Baade's Window (BW) at b= −4º as a sample of the Galactic bulge. Uncertainties that might arise in estimating a star's absolute magnitude due to distance spread along the line of sight through the bulge is small at this latitude (Frogel *et al.* 1990), only about ±0.2 mag. In the first part of this section we will derive an estimate for the ratio of luminous LPVs to non-variable giants in Baade's Window for direct comparison with our result for globular clusters. Then we will examine possible effects of [Fe/H] variations on this ratio.

---

[3]This reference value for $M_{bol}$ is chosen to facilitate comparison with Baade's Window; it corresponds to the faintest luminosity for which the Baade's Window luminosity function of Frogel & Whitford is well determined.

### 3.1  Counting Luminous LPVs in Baade's Window:
### A Comparison with Globular Clusters

Surveys for variable and non-variable red giants in BW have been carried out by: Lloyd Evans (1976) who found a total of 39 LLPVs in an area of 1200 arc min$^2$; by Blanco, McCarthy, & Blanco (1984) for stars of spectral type M6 or later in a 467 arc min$^2$ area; and by Blanco (1986) for stars of type M1 and later in a 42 arc min$^2$ area.  Frogel & Whitford's (1987) infrared photometry of an unbiased sample of the M giants found in the Blanco *et al.* surveys leads to the result that in the 467 arc min$^2$ BW field there are 1310 M stars with $M_{bol} \leq -1.2$[4]. DePoy *et al.* (1993) in their 2.2 μm survey of BW concluded that there are essentially no cool, luminous stars missing from the Blanco *et al.* grism surveys, i.e. *there are no luminous K giants in BW*.  Thus, the 1310 M stars with $M_{bol} \leq -1.2$ must be essentially the entire population of giants brighter than this limit in BW[5]. Lloyd Evans found 39 LLPVs in a search area 1200 arc min$^2$.  Of these, 15 are in the 467 arc min$^2$ covered by the Blanco *et al.* surveys of Baade's Window.  Thus the ratio of giants to LLPVs is 86 to 1, nearly 40% greater than the 62 to 1 ratio for globular clusters.

We can also compare the ratio of number of LLPVs to the integrated $K$ luminosity for globular clusters and the bulge.  In the previous section this ratio for the clusters was found to be 1.55 ($\times 10^{-4}$) for stars with $M_{bol} \leq -1.2$.  In BW the integrated $M_{Ko}$ for stars with $M_{bol} \leq -1.2$ is $-13.6$ (Frogel & Whitford [1987] with a distance of 8 kpc).  We would then predict 43 LLPVs in the 467 arc min$^2$ area searched for M stars whereas only 15 are observed.

### 3.2  The Effects of Metallicity on the Number of Luminous LPVs

In the previous sub-section we showed that there is a considerable difference in the frequency of occurrence of LLPVs relative to non-variable giants between Baade's Window and the metal rich globular clusters.  In the remainder of section 3 we will investigate the possibility that this difference is due to the difference in the metallicity distribution of the stars in the two populations.

One obvious effect of metallicity on the relative numbers of LLPV and non-LLPV giants is the absence of LLPVs from globular clusters with [Fe/H] < −1.0.  Thus, for a stellar population with an age comparable to those the globular clusters there is a minimum value of [Fe/H] required to produce an LLPV as defined above, i.e. with $M_{bol}$ brighter than the tip of the RGB.  What about stars with [Fe/H] in the range −1.0 to +1.0, i.e. that which encompasses the metal rich globulars of Table 1 and the stars in BW?

First, we examine the [Fe/H] distributions of BW stars and of the globular clusters that make up the sample of Table 1.  Table XI in Rich's study (1988; see also Whitford & Rich 1983) lists [Fe/H] for 100 K giants in Baade's Window.  This sample should be representative of the [Fe/H] distribution for all K giants in BW.  Re-examination of a subset of these 100 has led McWilliam & Rich (1994) to propose a downward revision by an average of −0.32 dex to the values in Rich's Table XI.  We have applied this correction to Rich's "solution 1", binned the resulting values, and list them in col. 2 of Table 3.  Values for the bins given in the first column of Table 3 are the *centers* of bins that are 0.25 dex wide.  To form a comparable distribution for the clusters in Table 1 we again used the approximation that the integrated *K* light of the clusters will be proportional to the relative numbers of stars in the clusters, binned and summed the values given in the last column of Table 1, and entered the resulting numbers in col. 3 of Table 3.  Figure 1 plots these relative distributions of giants in BW and the metal rich globulars of Table 1.  Not only do the BW stars extend to considerably higher values of [Fe/H] than the clusters, but even in the region of overlap the BW stars are weighted towards higher [Fe/H].  Thus, in order to do a proper comparison of relative numbers of LLPVs in BW and the clusters as a function of [Fe/H], we need to take into account the different [Fe/H] distributions.

---

[4] Frogel & Whitford used 7 kpc or $(m-M)_0$ of 14.2 to derive their luminosity function (their Table 8).  Since 8 kpc seems to be a more likely value for the distance (*e.g.* Reid 1989), the luminosity function used here for Baade's Window was recalculated from the Frogel & Whitford data to correspond to this revised distance.

[5] As an additional check on the possible presence of bright non-M giants that would not have been found in the Blanco *et al.* surveys, we selected all of the obviously bright stars on Blanco's (1986) chart of BW that were not classified as M type.  Nearly all of these have blue optical colors (Arp 1965) strongly implying that they are foreground stars.  Most of the few remaining ones not observed by Arp and not selected in the Blanco *et al.* surveys are crowded.

Figure 2 compares the [Fe/H] distributions of the LLPVs and the giants in the 12 clusters of Table 1.  The [Fe/H] distribution for the giants is determined as described in the previous paragraph.  Note the relative scaling of the left and right axes to make the visual comparison of the two distributions easier.  This figure shows that to within the uncertainties arising from the small number of clusters and LLPVs, the relative number of LLPVs to giants in the clusters is independent of [Fe/H].

We now renormalize the distribution over [Fe/H] of the cluster LLPVs to correspond to the [Fe/H] distribution of the giants in BW.  To do this we assume that in each [Fe/H] bin for the clusters the number of LLPVs is given by a constant times the number of giants.  It is not necessary to assume that the constant is the same for all bins, although this does appear to be approximately true.  It is also not necessary to assume that the constants are the same for BW since we will redistribute the cluster giants in the [Fe/H] bins while keeping their total number constant and then redistribute the cluster LLPVs according to the new giant distribution.  The multiplicative factors for cluster stars for each [Fe/H] bin are listed in col. 6 of Table 3.  These values are a constant times those that would obtain if one just took the ratio of BW to globular cluster stars in each bin.   The new [Fe/H] distribution for cluster LLPVs is given in col. 7.  Note that it was not necessary to explicitly make the new total number of LLPVs equal the old number.  The near equality of the two numbers is a result of the fact that for the 3 most populous bins the numbers of LLPVs per giant are close to one another.  Figure 3 compares this renormalized distribution in [Fe/H]  of cluster LLPVs with the [Fe/H] distribution for stars in BW.  Once more, within the uncertainties, the two distributions appear to be identical for the 4 bins of overlap.  This result follows directly from the fact that the ratio of LLPVs to giants in the clusters is very similar for all [Fe/H] bins.

The conclusion from the above is that over the [Fe/H] range in which *cluster* LLPVs (as defined earlier) are found, their distribution relative to non-LLPV giants both in the clusters themselves and in the bulge as represented by BW is independent of metallicity.  We must now determine whether this ratio of LLPV to non-LLPV giants is the same for the *bulge* LLPVs themselves.  In order to do this we need to devise a way of estimating [Fe/H] for the BW LLPVs.

### 3.3  [Fe/H] for the LLPVs in Baade's Window Based on their Periods

Several of the physical properties of globular cluster variable stars correlate with the parent cluster's metallicity (*e.g.* Feast 1981, 1996; Lloyd Evans 1983; and Menzies and Whitelock 1985). In particular, Feast (1981) and Lloyd Evans (1983) have shown that the periods of the cluster LPVs are closely correlated with [Fe/H]. For metal-poor clusters, they include SRd variables which lie at the top of their parent cluster giant branches and appear to be the metal poor analogues of the LLPVs found in clusters such as those listed in Table 1 here (*e.g.* Whitelock 1986; Feast 1996). In order to estimate [Fe/H] for Baade's Window LLPVs we will quantify the relation between log P and [Fe/H] for globular clusters and then assume that it is applicable to the BW stars.

A physical rational for a relationship between log P and [Fe/H] can be inferred from the pulsation equation. From Glass and Feast (1982) we have

$$\log(Q \mathcal{M}^{-1/2}) = \log P + 0.3 \times M_{bol} + 3 \times \log T_{eff} - 12.71 \qquad 3.$$

where Q is the pulsation constant, $\mathcal{M}$ the stellar mass, and $M_{bol}$ the mean bolometric magnitude. Giant stars in high metallicity clusters have lower effective temperatures at the same luminosity than giants in clusters of lower metallicity (as well as having a slightly larger mass for the same age). Giants in a higher metallicity cluster will attain higher luminosities (more negative $M_{bol}$) than ones in lower metallicity clusters because of the linear relation between core mass and luminosity on the AGB and the inverse relation between evolutionary rate and metallicity (Iben & Renzini 1983 and references therein). Hence, it is reasonable to infer the existence of a log P - [Fe/H] relation for LLPVs.

To quantify the log P, [Fe/H] relation for globular cluster variables we will consider only those clusters with [Fe/H] > −1.3, the lower limit of [Fe/H] for the K giants in BW (McWilliam & Rich 1994). From this sample of clusters we will limit the selection of variable stars as follows: First, we select those variables that lie close to or above the top of the RGB; this includes all of the LLPVs discussed in this paper. If a cluster has one or more LLPVs in it, no further variables from that cluster will be included in the solution. This leads to the exclusion of V2 and V8 from NGC6712 and V8 from 47 Tuc. For clusters without LLPVs (mainly the ones near the low end of our allowed [Fe/H] range) we include the variables that lie at the tops of the cluster RGBs; these are the SR stars, the suggested metal-poor analogues of LLPVs. Specifically, we include NGC288 - V1; NGC362 - V2; NGC1851 - V24; and NGC6121 - V4. With these stars we find

$$[Fe/H] = 1.80 \log P - 4.81 \qquad 4.$$

with an uncertainty of ±15% in the predicted value of [Fe/H] based on the scatter in log P. For clusters with more than one LLPV the range in log P for them is small but real and must arise from small differences in the evolutionary state of the different stars. Thus stellar evolution alone, most likely the non-zero spread in the mass of stars becoming LPVs, imposes a limit to the accuracy of this equation. Various, arguably justifiable, changes in the sample selection do not significantly alter this result (*e.g.* not including the 4 SR stars listed above). This equation can now be applied to bulge LLPVs with the resulting [Fe/H] distribution given in col. 4 of Table 3.

Figure 5a shows the period distributions for the globular cluster LLPVs (Table 1) and for all of the LLPVs found in BW by Lloyd Evans (1976). Figure 5b is the same except now the numbers for the cluster LLPVs have been normalized to correspond to the [Fe/H] distribution in BW. In other words, these are the numbers of cluster LLPVs that would be observed if the [Fe/H] distribution of the clusters were the same as the [Fe/H] distribution of the BW M giants. This procedure is described above with normalization factors given in col. 7 of Table 3. We emphasize that this normalization was derived from the non-LLPV giants alone. Fig. 5b shows the close similarity of the period distributions of the two stellar populations in the 3 bins of overlap once differences due to different [Fe/H] distributions have been taken into account. It also draws attention to the fact that 15 of the 33 bulge LLPVs have periods longer than the longest period for a globular cluster LLPV, although the longest period is only 470 days which is just 0.18 dex larger than the longest period globular LLPV. Therefore, although it is necessary to extrapolate the above equation to apply it to all of the bulge LLPVs observed by Glass & Feast (1982) and Frogel & Whitford (1987), the extent of the extrapolation is only by 0.2 dex.

An important caveat to this approach is that if bulge LLPVs are significantly younger, i.e. more massive, than the globular cluster LLPVs *of the same [Fe/H]*, the relation between log P and [Fe/H] could be different for the two groups of stars as can be seen from eqn. 3. Available evidence, though, points to the similarity of cluster and bulge LLPVs, at least for ones of similar period. First, we note that bulge LLPVs obey the same P-L relation as do those in globular clusters and the LMC (Glass and Feast 1982, Feast 1984 and references therein, Feast 1996). Also, four of the shorter period (P ~ 230d) bulge LLPVs have been investigated specifically to determine their spectral types at maximum light; these types were found to lie on a standard (for globular clusters) period-spectral type relation for LLPVs at maximum light (Feast 1972).

Another piece of evidence in favor of a comparable age for the bulge and cluster LLPVs is that the $M_{bol}$ distributions for both are not statistically different from one another (Frogel *et al.* 1990, Fig. 13). Figure 6 compares the two distributions of LLPVs over $M_{bol}$. The cluster LLPV distribution has been normalized to correspond to the Baade's Window [Fe/H] distribution as described earlier. Note that the fact that the 2 most luminous Baade's Window LLPVs have longer periods than any found in the clusters is not inconsistent with the pulsation equation and our line of reasoning. Thus, with no strong evidence to the contrary, we will conclude (as originally suggested by Lloyd Evans 1976) that the presence in BW of stars with a higher metal content than any found in the globular clusters with known LLPVs is the main reason that the LLPV population of BW contains stars of longer period than any found in the clusters.

Figures 7 and 8 illustrate the main result of this section - the derived [Fe/H] distribution of the BW LLPVs. Fig. 7 compares this distribution to that for globular cluster LLPVs, while Fig. 8 compares it to that for the K giants in BW. In spite of the fact that the frequency distributions of bulge and cluster giants are similar at the low [Fe/H] end (Fig. 3), Fig. 7 shows an absence of low [Fe/H] LLPVs from Baade's Window (i.e. an absence of ones with periods as short as those found in the lower metallicity clusters). Not surprisingly, Fig. 7 also shows that Baade's Window has a significant number of LLPVs with a somewhat higher metallicity than those found in clusters.

Figure 2 showed that the [Fe/H] distributions of LLPVs and non-LLPV giants from clusters are similar. This is clearly not the case for BW as may be seen from Figure 8. In particular, note the absence of any LLPVs with [Fe/H] >0.0. If their frequency of occurrence relative to non-LLPV giants were the same as in the bins with [Fe/H] between 0.0 and −0.75, one would predict that there should be two dozen LLPVs in the bulge with [Fe/H] >0.0 (i.e. periods greater than 470 days) whereas none are found.

We note that the close similarity of the $M_{bol}$ distribution of cluster and bulge LLPVs (Fig. 6) contrasts with the presence of a number of LLPVs in the bulge with longer periods than those in the clusters. This raises the possibility that the P-L relation is not single valued, but also depends on some combination of [Fe/H] and mass, contrary to our discussion above.

In the next sub-section we will discuss the hypothesis that no metal rich LLPVs are observed in the bulge because their lifetimes are too short. Here we consider, and reject, the possibility that they have been missed because of observational limitations. The Blanco *et al.* surveys of BW for M giants and Lloyd Evans' (1976) search for variables were carried out in wavelength regimes where blanketing of the spectrum, especially by TiO and VO, becomes more and more severe in cooler and more metal rich giants. Nonetheless, for both surveys the faintest cool stars found were still considerably brighter than the limits of detectability for the surveys. Furthermore, out of two dozen predicted, but missing, LLPVs, one would expect at least a few to have been near maximum light and hence maximum temperature at the times the surveys were carried out. We can also rule out the possibility that longer period LLPVs are so bright that they would have been missed by Blanco's grism surveys due to saturation since DePoy *et al.* (1993) pointed out that essentially all of the stars on Blanco's direct and grism plates that were too bright to classify correspond to blue, presumably foreground, stars. Neither would they have been missed by Lloyd Evans' survey.

Perhaps the most compelling argument against the possible presence of more than just a very small number of as yet undetected long period, hence luminous and red, LLPVs in Baade's Window is the limit set by IRAS (Frogel & Whitford 1987; section 5.1 of Frogel 1988, 1998). There is no question that quite a few IRAS sources detected in the bulge are previously unknown LLPVs with especially long periods (*e.g.* Whitelock, Feast, & Catchpole 1991). However, in the area of BW surveyed *optically* for LLPVs, 11 of the 12 IRAS sources were previously identified as LLPVs by Lloyd Evans (1976). Only at lower latitudes does the fraction of IRAS sources previously noted as LLPVs drop (Feast 1986). At these lower latitudes, though, optical detection becomes more and more difficult.

### 3.4  The Missing High [Fe/H] Bulge LLPVs - Rapid Evolution?

There is considerable observational evidence (Frogel and Elias 1987, Whitelock, Pottasch, & Feast 1986) to support the theoretical contention (e.g. Wood 1979; Jones, Ney, & Stein 1981) that the energy supplied by the pulsations of a large amplitude variable will significantly enhance any other mass-loss mechanism such as radiation pressure on grains (e.g. Jura 1984; Knapp 1986). If this is the case for the bulge LLPVs, we should then see a correlation between pulsation period - which is proportional to amplitude for LLPVs (Feast *et al.* 1982; Whitelock & Feast 1993) - and some measure of the extent of a star's circumstellar shell (CSS). Frogel and Whitford (1987) demonstrated that the LLPVs in BW are markedly redder in $K$-$L$ at any given $J$−$K$ than are non-LLPV M giants. This redness is due to excess emission at $L$ rather than relative faintness at $K$. The excess emission may be plausibly attributed to thermal radiation by dust in the circumstellar environment, particularly given the correlation between $K$−$L$ and K-[10] (Frogel & Whitford 1987). In Figure 9a we plot $K$−$L$ versus log P for the bulge and cluster LLPVs.[6] For comparison we also show the locations of the non-LLPV cluster variables. For clarity we use a different symbol for the BW LLPVs that have periods longer than the longest period possessed by a cluster LLPV. It is clear from this figure that the greatest excess emission is associated with the longest period BW LLPVs. Figure 9b shows that with the exception of a few with longest period and greatest excess, the BW LLPVs have an $M_{bol}$ distribution which overlaps that of the globular cluster LLPVs (see also Fig. 6).[7] Thus we have demonstrated the existence of the sought after correlation between pulsation period and extent of the circumstellar envelope.

---

[6]Colors and $M_{bol}$ for the Baade's Window variables are from Frogel and Whitford (1987) or from Glass and Feast (1982), after conversion to the CIT/CTIO photometric system, if not given in the former. TLE136 and 395 (Lloyd Evans 1976) were not included as LLPVs since Lloyd Evans considers them to be semi-regular variables rather than Miras. For D9 data from Glass and Feast were used rather than from Frogel and Whitford because of possible mis-identification. For TLE 181 $M_{bol}$ was recalculated with an average of the photometry from the two sources.

Based on models of Bowen & Willson (1991), Willson, Bowen, & Struck (1995) have shown that an increase in [Fe/H] alone leads to a rapid increase in grain formation and hence in the mass loss rate for Mira variables. Since period, pulsation amplitude, and [Fe/H] are all positively correlated in Miras, or LLPVs, from an old population, their result and the discussion in the previous paragraph are consistent with one another. Clearly, in an old population, [Fe/H] is the fundamental physical parameter that is responsible for other aspects of a star's evolution and appearance, including its pulsation properties.

Examination of the $K$–[10] color (Fig. 9 of Frogel and Whitford 1987) of the BW LLPVs indicates that the range of the excess emission is at least 3 mags - a factor of 16. The circumstellar emission is most likely optically thin which means that it should be proportional to the amount of material in the CSS and hence some combination of the mass loss rate and the metallicity; dependence on other parameters will be relatively less important (cf. Frogel and Whitford 1987, Frogel and Elias 1987). Based on our (small) extrapolation of the log P, [Fe/H] relation determined for globular cluster LLPVs, the range in [Fe/H] for the longer period LLPVs in Baade's Window is a factor of 2 to 3; the estimated range in mass-loss rates and extent of the CSSs would then have a range of 2 to 4 times greater than the [Fe/H] range. These higher mass-loss rates result from a combination of more energetic pulsations, more grain formation, and more molecular absorption, particularly from $H_2O$, in the outer stellar atmosphere. All of these effects ultimately arise from higher [Fe/H].

Higher mass loss rates imply significantly shorter lifetimes for LLPVs. The lifetime of an LLPV will be determined by how long the stellar envelope lasts (cf. Iben and Renzini 1983). Thus, to first order lifetimes will be inversely proportional to mass loss rate (differences in envelope mass as a function of [Fe/H] are assumed to be small). Since their envelopes are being rapidly blown away, their evolution as AGB stars will be quickly terminated. We point out, based on the data in Frogel & Whitford (1987) that the expected 10 μm flux from many of these stars would be too low for them to have been detected by IRAS. This is consistent with Feast's (1986) finding that only between one-third and one-quarter of the known LLPVs in the Baade's Window field were detected by IRAS. The lower latitude "windows" with more longer period stars also have a high rate of detection by IRAS (Feast 1986), although it is difficult to judge how complete the optical surveys for LLPVs are in these very crowded and reddened fields.

Thus, although our evidence is somewhat circumstantial and arguments rather qualitative, we conclude that the most likely explanation for the relative scarcity of high [Fe/H] LLPVs in Baade's Window is due to lifetimes for these stars that are significantly shorter than the lifetimes of LLPVs with lower [Fe/H]. These abbreviated lifetimes are, in turn, due to metallicity enhanced mass loss rates that may set in rather abruptly near [Fe/H] = 0.0 (cf.

---

[7] The number of observations per BW LLPV is not large, but the data in FW and Glass & Feast (1982) indicate that the values plotted are not too far from the mean values. In particular, note that none of the longest period LLPVs have $M_{bol}$ fainter than the mean for the other LLPVs.

Willson *et al.* 1995). In our qualitative argument we have neglected any possible effect of mass loss on the first giant branch. These, too, could be metallicity dependent.

Frogel (1988), Frogel *et al.* (1990), and Whitelock & Feast (1993) analyze the surface brightness and surface density distributions along the minor axis of the bulge of several sources of extended emission and of stellar components. They find that the falloff in surface density of the LLPVs is similar to that of all M giants or of the integrated K light of the bulge but less steep than that of the latest M giants (M7-9). Since these latest M giants are tracers of the most metal rich component of the bulge's stellar population, and the presence of an [Fe/H] gradient in the bulge is well established (*e.g.* Tiede, Frogel, & Terndrup 1995) this result is consistent with our conclusion that LLPVs produced by the most metal rich part of the population will have significantly shorter lifetimes than those with somewhat lower metallicities.

### 3.5  The Comparison Between the Frequency of LLPVs in Clusters and the Bulge Revisited

Can we redo the comparison between Baade's Window and globular clusters from section 3.1 and recompute the frequency of occurrence of LLPVs in Baade's Window excluding the non-variables with [Fe/H] $\gtrsim 0.0$? Baade's Window contains a significant number of non-variable giants that are more luminous ($M_{bol} \leq -3.7$) than the brightest such giants that are found in clusters. Frogel & Whitford (1987) attributed the presence of such bright non-variables in an old population to the presence of stars that have metallicities greater than solar. Their evolutionary rate would be sufficiently slow compared to stars of sub-solar metallicity that on the AGB their envelope mass would be sufficient to allow them to evolve well above the point of core He flash on the RGB[8]. We have just argued that Baade's Window does not appear to contain any (or certainly no significant number of) LLPVs with [Fe/H] $\gtrsim 0.0$. If the number and integrated luminosity of non-variable M giants with $M_{bol} \leq -1.2$ were adjusted to include only stars with [Fe/H] $\leq 0.0$, to correspond approximately to what obtains for metal rich globular clusters, the two comparisons we just made would show considerably closer agreement between clusters and Baade's Window.

Consider first the ratio of numbers of LLPVs and non-variable giants. From the BW luminosity function (Frogel & Whitford 1987) we estimate that there are 110 stars with $M_{bol} \leq -1.2$ out of the 1310 M giants[9]. How many precursors would such luminous stars have? The [Fe/H] distribution of McWilliam & Rich (1994) shows that 38% of the Baade's Window stars have [Fe/H] $\geq 0.0$. For the purpose of this estimation, we assume that the luminosity function of giant stars is roughly independent of [Fe/H] in the range of interest. Then we have $0.62 \times (1310 - 110) = 744$. With 15 LLPVs in Baade's Window, we get a ratio of 50 to 1 for non-variable giants vs. LLPVs, certainly closer to the 62 to 1 ratio for globular clusters than found in section 3.1 and within the uncertainties of the small numbers involved.

---

[8] An alternative is that these luminous stars are several Gyr younger than the rest of the population *e.g.* McWilliam & Rich 1994. As such they would be analogues of luminous AGB stars seen in LMC clusters with ages of a few Gyr.

[9] Note that this approach is somewhat of an over simplification because of the small but real distance spread along the line of sight of BW stars. Because of their relative rarity though, it is unlikely that the sample of late M giants is contaminated by nearby field stars (see also Blanco *et al.* 1984).

Next consider the number of LLPVs relative to the integrated K luminosity of Baade's Window. From the integrated $M_{K_0}$ we first subtract the contribution from stars with $M_{bol} \leq -3.7$. Table 8 of Frogel and Whitford (1987) shows that although such luminous stars are few in number, they contribute significantly to the integrated $M_{K_0}$. The correction to the integrated $M_{K_0}$ is +0.47 mags. This new value of −13.1 is further reduced by 38% to allow for the progenitors of the bright stars. The adjusted integrated $M_{K_0}$ for stars with $M_{bol} \leq -1.2$ is now −12.6, which leads to a prediction of 17 LLPVs in BW, close to the 15 that are actually observed.

## 4. DISCUSSION AND CONCLUSIONS

Bolometrically and in the near-IR, long period variables will be the brightest stars in an old population. They most likely are located at the top of the AGB and mark the end of the nuclear burning lifetime for stars between one and a few solar masses. Studying them in an environment that allows determination of important physical characteristics such as luminosity and metallicity is essential for understanding their formation, evolution, and demise. Globular clusters present one such environment. It is unfortunate that information on the presence of LLPVs exists for only one-quarter of the known metal rich clusters. We hope that the recent spate of observations of heavily reddened metal rich globular clusters with modern array detectors will lead to a significant increase in size of the present sample of cluster LLPVs. Still, we are able to draw tentative conclusions based on the small existing sample and make a comparison with LLPVs in the Galactic Bulge.

First of all, we find that for globular clusters with [Fe/H] $\gtrsim$ −1.0 LLPVs occur with a frequency per non-variable giant that appears to be independent of [Fe/H]. To say any more about an [Fe/H] dependence would require doubling the sample of known globular LLPVs. With this result, we can apply the formalism developed by Renzini (1994) based on the fuel consumption theorem (Renzini 1981) to the sample of globular clusters in Table 1 and estimate the lifetime of a typical cluster LLPV. Rewriting Renzini's (1994) eq. 1.10 slightly, we have

$$N_{LPV} = B(t) L_{bol,GC} \tau_{LPV} \qquad 5.$$

where $N_{LLPV}$ and $\tau_{LLPV}$ are the number of LLPVs in a stellar population and their lifetime, respectively, $B(t)$ is the specific evolutionary flux of the stellar population (measured in stars yr$^{-1}$ $L_\odot^{-1}$ ), a quantity that is nearly insensitive to choice of the IMF and mass loss rate and which depends only weakly on age, and $L_{bol,GC}$ is the total bolometric luminosity of the stellar population. If we take for the stellar population the sample of clusters in Table 1, then $N_{LLPV}$ = 19, $L_{bol,GC}$ is = $2.9 \times 10^6$ L$_\odot$ (assuming a typical bolometric correction to the $K$ magnitude of a cluster of 2.2), and from Renzini's (1994) figure 1.3 $B(t) = 2.2 \times 10^{-11}$ , we arrive at a typical lifetime for a cluster LLPV of $3 \times 10^5$ yr., in agreement with the lifetime estimate of Renzini & Greggio (1990) based just on the 4 LLPVs in 47 Tucanae. In this regard it is interesting to note the following: From the models of Sweigart & Gross (1978), it takes about 200 Myr for a star with $\mathcal{M}_{ZAMS}$ between 0.7 and 1.1 $\mathcal{M}_\odot$ to evolve up the giant branch from log (L/L$_\odot$) = 1.10 to the onset of core He flash in the [Fe/H] range of relevance for this study. Quick inspection of the existing cmds of the dozen clusters in Table 1 suggests that they have a total of 10,000 stars in this luminosity range. Since these same clusters only have 19 LLPVs, the implied lifetime for an LLPV is $4 \times 10^5$ yr.

The galactic bulge is an environment that is also well suited to the determination of physical characteristics of LLPVs because its distance is known and surveys of some areas yield complete samples of LLPVs. The frequency of occurrence of LLPVs in such a population will also be of considerable relevance for the interpretation of potential pixel lensing events as may be seen in the bulge of other galaxies. For Baade's Window in particular, optical surveys for LLPVs are believed to be nearly complete. Near-IR data exist for these stars which permits detailed comparisons with their globular cluster counterparts.

We have had to estimate [Fe/H] for the BW LLPVs based on a log P, [Fe/H] relation derived from globular cluster stars. We find that in the [Fe/H] range in common to both BW and globular clusters the frequency distribution of LLPVs is the same in both populations. In other words, the rate of production of LLPVs per non-LLPV is the same in clusters and in Baade's Window. Furthermore, this rate appears to be independent of [Fe/H] in the range $-1.0 \leq$ [Fe/H] $\leq -0.25$. Greater than this metallicity, however, the BW field appears to be markedly deficient in LLPVs, i.e. it has far fewer long period LLPVs that one would expect based on the [Fe/H] distribution of non-variable giants.

We argue that this deficiency of metal rich LLPVs is not due to problems with detecting such stars. For example, take as an alternative to truncated lifetimes for metal rich LLPVs the possibility that the most metal rich AGB stars do not go through an optically detectable LLPV phase but quickly become OH/IR stars. The OH/IR sources plausibly represent an extension of the LLPV sequence to the longer periods and higher luminosities that would be expected for more metal rich and/or higher mass stars (Feast 1985, Whitelock 1996). Although Lloyd Evans would not necessarily have detected OH/IR sources in BW, IRAS would have had they existed in any significant numbers. Nearly all IRAS sources in this area, though, can be identified with known stars most of which are LLPVs (Feast 1986, Frogel & Whitford 1987).

To explain the relative absence of long period, high [Fe/H] LLPVs in Baade's Window we propose that their lifetimes are significantly curtailed by virtue of their having mass loss rates several times greater than those of less metal rich stars. Such higher mass loss rates can result from a variety of factors all of which are generally [Fe/H] related: *e.g.* larger amplitude pulsations; greater formation rate of grains, hence greater radiation pressure driven mass loss; greater radiation pressure on [Fe/H] sensitive molecules in the outer atmosphere. A point in favor of this argument is the correlation shown by the LLPVs between their period and evidence for a circumstellar envelope (Fig. 9). A strong test of this hypothesis awaits an enlarged sample of globular cluster LLPVs. Although a significant fraction of the stars in BW are considerable more metal rich than any cluster, one can hope that any trend of LLPV lifetime with [Fe/H] will begin to reveal itself amongst the cluster population.

Finally, Figs. 6 and 9 demonstrate that with the exception of a few of the Baade's Window LLPVs with the longest periods and greatest $K-L$ excess, the distribution in $M_{bol}$ overlaps that of the cluster LLPVs. Hence it is incorrect to say , as already pointed out by Frogel (1988) and Renzini & Greggio (1990), that an old population such as is found in the bulge cannot produce luminous LLPVs.

We thank Dr. C. Clement for giving us a computer readable version of the most recent update of the late H. S. Hogg's catalogue of variable stars in globular clusters (Hogg 1973) and Alvio Renzini for illuminating conversations. JAF's research on globular clusters is supported in part by NSF grant AST-9218281. He also acknowledges the hospitality of the SAAO and its staff during a visit when the two authors first developed some of the ideas in this paper, and a Visiting Senior Research Fellowship at the University of Durham during which time those ideas were brought to fruition. Both JAF and PAW thank ESO for supporting visits by them to Garching which facilitated the completion of this paper.

# FIGURE CAPTIONS

Figure 1 − The relative distribution over [Fe/H] of giants in Baade's Window and in the sample of metal rich globular clusters that have been searched for LLPVs. Numbers are from Table 3. The scaling of the left and right axes has been chosen to emphasize the much greater relative number of metal rich stars in the Baade's Window distribution.

Figure 2 − The distribution over [Fe/H] of globular cluster giants and LLPVs for the clusters of Table 1. The values are from Table 3. The axes have been scaled to emphasize the similarity of the two distributions.

Figure 3 − The relative distribution over [Fe/H] of giants from Baade's Window and LLPVs from globular clusters. The latter distribution has been normalized as described in the text. Numbers are from Table 3. The left and right axes have been scaled to emphasize the similarity of the two distributions in the [Fe/H] region of overlap.

Figure 4 − Log P versus [Fe/H] for all globular cluster variables on the RGB with near-IR observations from sources noted in the text. The relation between log P and [Fe/H] given in the text (shown by the straight line) is derived from the 15 LLPVs and 4 luminous SR variables denoted by the filled triangles in the figure (note that there are two pairs of LLPVs - from NGC6356 and 6637 - with nearly identical periods so that only 13 filled triangles are actually visible). The vertical arrow at [Fe/H] = −1.3 marks the lower limit of clusters considered.

Figure 5 − a) The observed distributions of log (period) for globular cluster LLPVs from Table 1 of this paper and all of the LLPVs found in Baade's Window by Lloyd Evans (1976). b) This figure shows the globular cluster LLPV period distribution after the numbers of LLPVs have been normalized to correspond to the [Fe/H] distribution observed in Baade's Window as described in the text.

Figure 6 − The distribution in $M_{bol}$ of globular cluster and BW LLPVs. For BW the magnitudes are taken from Frogel & Whitford (1987) and Glass & Feast (1982), the latter after conversion to the CIT/CTIO system. A distance of 8 kpc to the Galactic Center is used. The cluster LLPV distribution has been normalized to correspond to the [Fe/H] distribution observed in BW as described in the text. The brightest BW LLPV actually has an $M_{bol}$ of −6.1.

Figure 7 − [Fe/H] values for the Baade's Window LLPVs have been estimated from their periods as describe in the text. The [Fe/H] distribution of cluster LLPVs has been normalized as in previous figures and is described in the test.

Figure 8 − [Fe/H] values for the BW LLPVs have been estimated from their periods as described in the text. The [Fe/H] distribution for BW giants is from McWilliam & Rich (1984).

Figure 9 − a) $(K−L)_0$ colors for LLPVs in globular clusters and the bulge are plotted against log P. Sources for these data are given in the text. LLPVs from the bulge with periods longer than 310 days, the upper limit to the period of cluster variables, are plotted with a different symbol. Non-LLPV variables from clusters are also plotted for comparison purposes. b) Same data set as in (a) but now for $M_{bol}$ versus log P. Note the rather small range occupied by both the cluster and bulge LLPVs except for two of the stars with the largest $(K−L)_0$ color.

**Table 1. Metal-rich clusters with LLPV content known**

| NGC | [Fe/H] | $M_{V,0}$ | $(V-K)_0$ | $M_K$ | #LLPVs | $10^{-(0.4 \times M_{K_0})-3}$ |
|---|---|---|---|---|---|---|
| 104 | −0.76 | −9.26 | 2.78 | −12.04 | 4 | 65.6 |
| 5927 | −0.37 | −7.66 | 3.00 | −10.66 | 1 | 18.4 |
| 6171 | −1.04 | −6.98 | 2.77 | −9.75 | 0 | 8.0 |
| 6352 | −0.70 | −6.31 | 2.98 | −9.29 | 1 | 5.2 |
| 6356 | −0.50 | −8.35 | 2.78 | −11.18 | 4 | 28.2 |
| 6362 | −1.06 | −6.82 | 2.54 | −9.36 | 0 | 5.6 |
| 6388 | −0.60 | −9.65 | 2.54 | −12.19 | 3 | 75.0 |
| 6553 | −0.25 | −7.59 | 3.01 | −10.60 | 2 | 17.3 |
| 6637 | −0.71 | −7.35 | 2.64 | −9.99 | 2 | 9.9 |
| 6712 | −1.01 | −7.35 | 2.49 | −9.84 | 1 | 8.6 |
| 6723 | −1.12 | −7.67 | 2.50 | −10.17 | 0 | 11.7 |
| 6838 | −0.73 | −5.40 | 2.69 | −8.09 | 1 | 1.7 |

**Table 2. Details of the Z=0.01 model**

| s | $V-K$ | $M_{K_0}$ | %K | $M_{K_0}$ $(M_{bol} \leq -1.2)$ | LLPVs | #giants | %AGB | GB/LLPV |
|---|---|---|---|---|---|---|---|---|
| (1) | (2) | (3) | (4) | (5) | (6) | (7) | (8) | (9) |
| 0.0 | 2.89 | -6.20 | 54 | -5.53 | 0.022 | 1.57 | 31 | 71 |
| 2.35 | 2.93 | -5.77 | 41 | -4.80 | 0.015 | 0.78 | 31 | 52 |

**Table 3. Numbers of giants and LLPVs in Baade's Window and globular clusters (GCs)**

| [Fe/H] | Relative # of Giants | | # of LLPVs (observed) | | normal. factor | GC LLPVs |
|---|---|---|---|---|---|---|
| | BW (observed) | GCs from $M_K$ | BW | GCs | for GC LLPVs | [Fe/H] normal. |
| (1) | (2) | (3) | (4) | (5) | (6) | (7) |
| -1.38 | 2 | | 0 | | | |
| -1.13 | 7 | 34 | 0 | 1 | 0.53 | 1.2 |
| -0.88 | 5 | 66 | 0 | 4 | 0.19 | 1.8 |
| -0.63 | 14 | 120 | 7 | 11 | 0.30 | 7.4 |
| -0.38 | 18 | 36 | 14 | 3 | 1.29 | 8.8 |
| -0.13 | 16 | | 12 | | | |
| 0.13 | 21 | | 0 | | | |
| 0.38 | 15 | | 0 | | | |
| 0.63 | 2 | | 0 | | | |
| 0.88 | 0 | | 0 | | | |

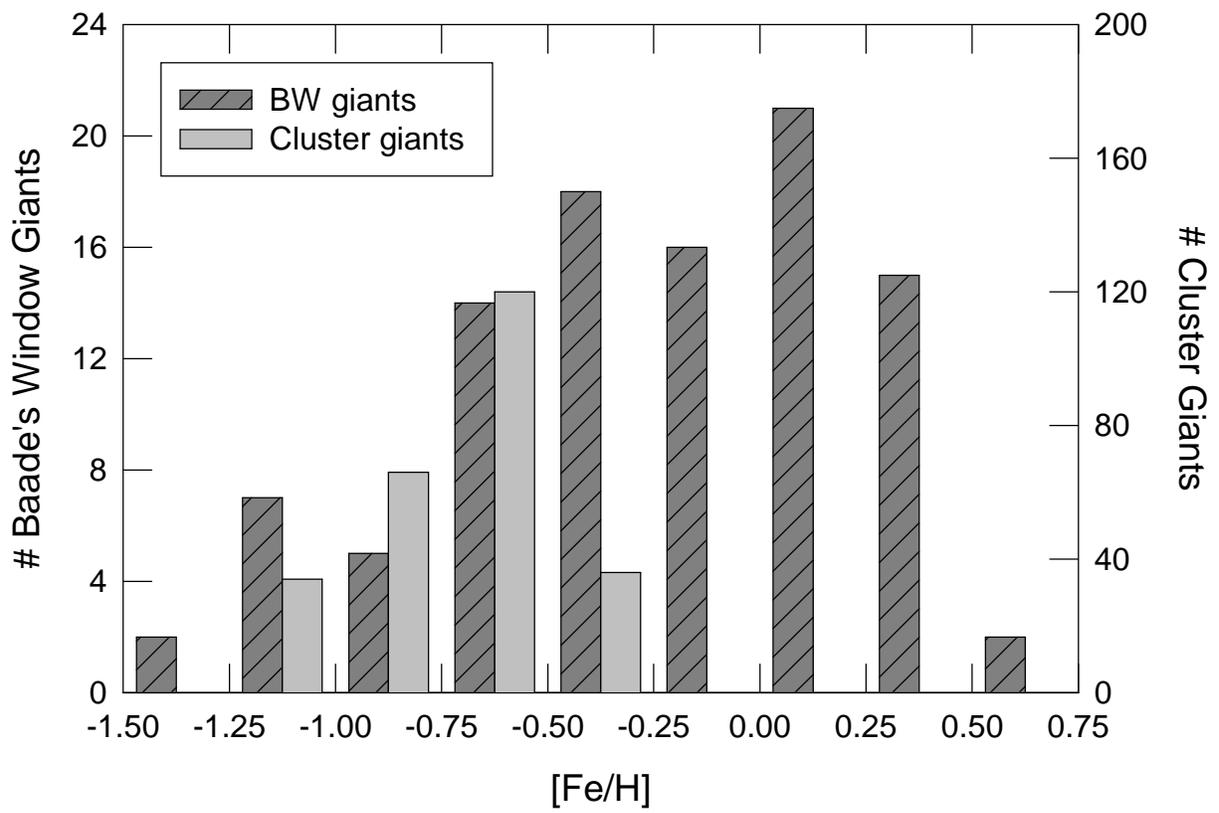

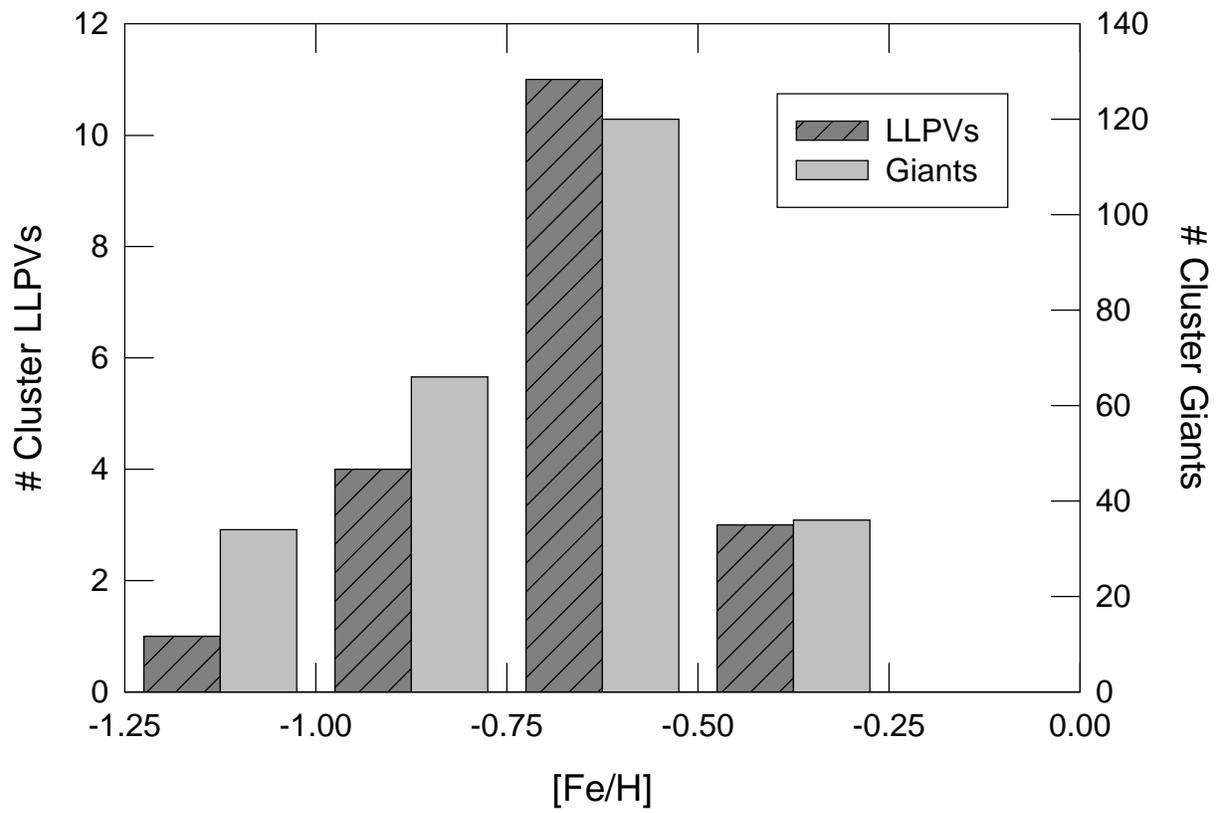

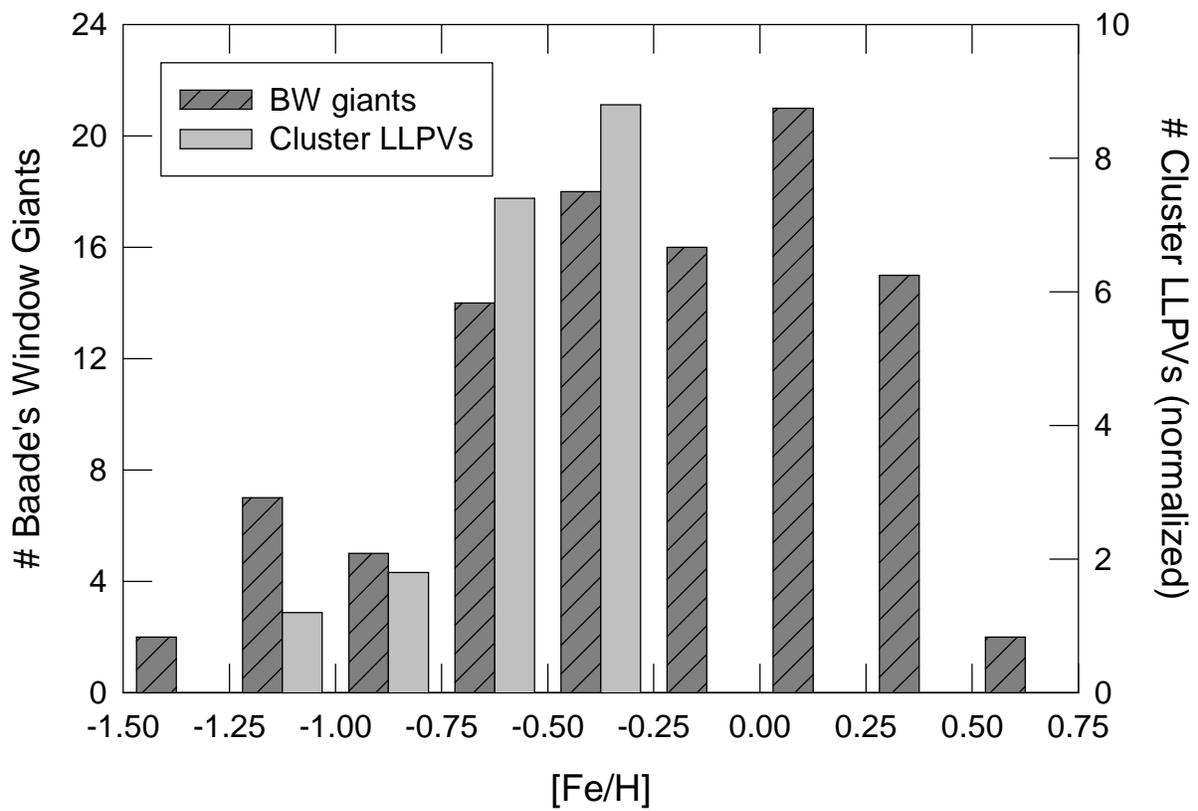

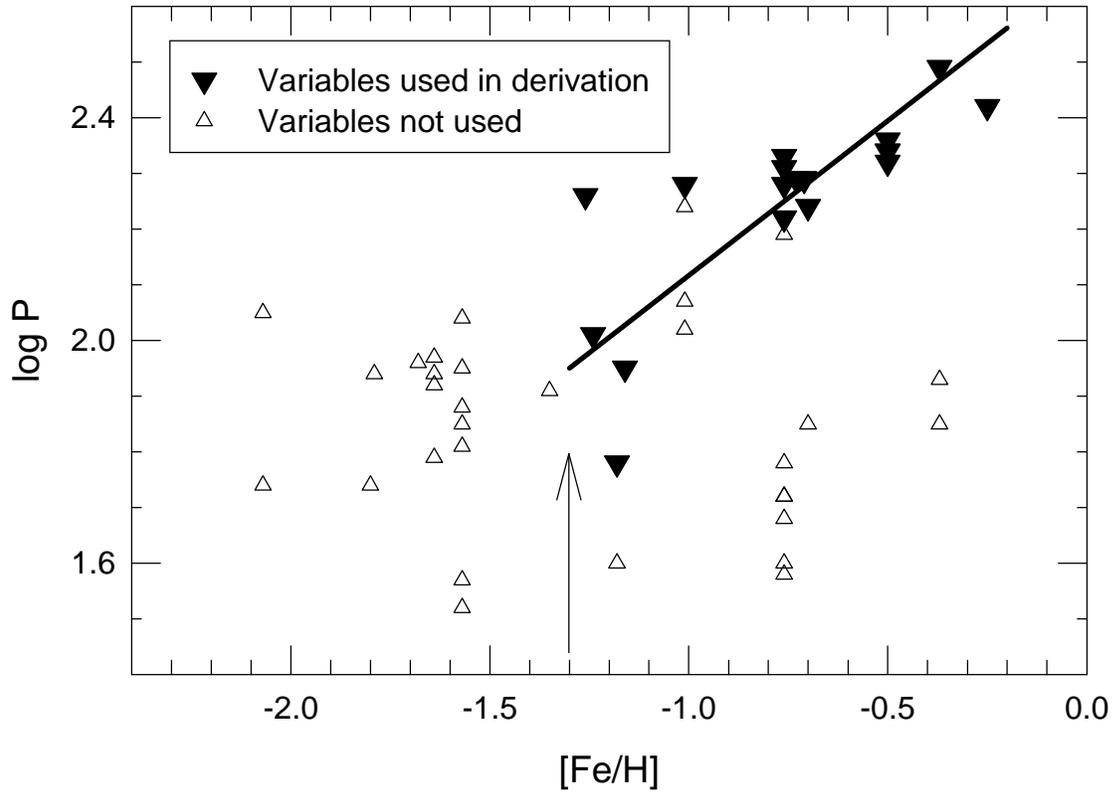

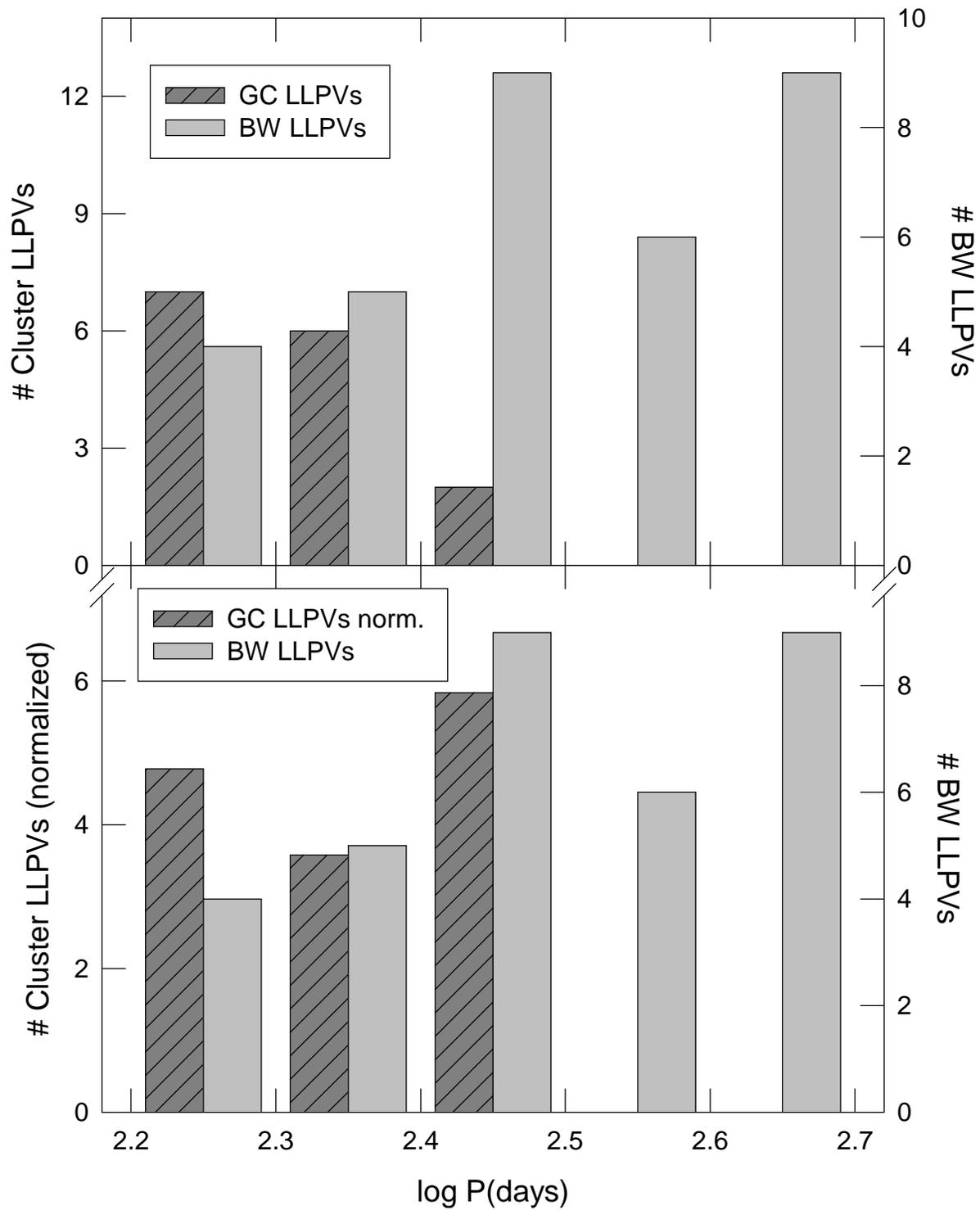

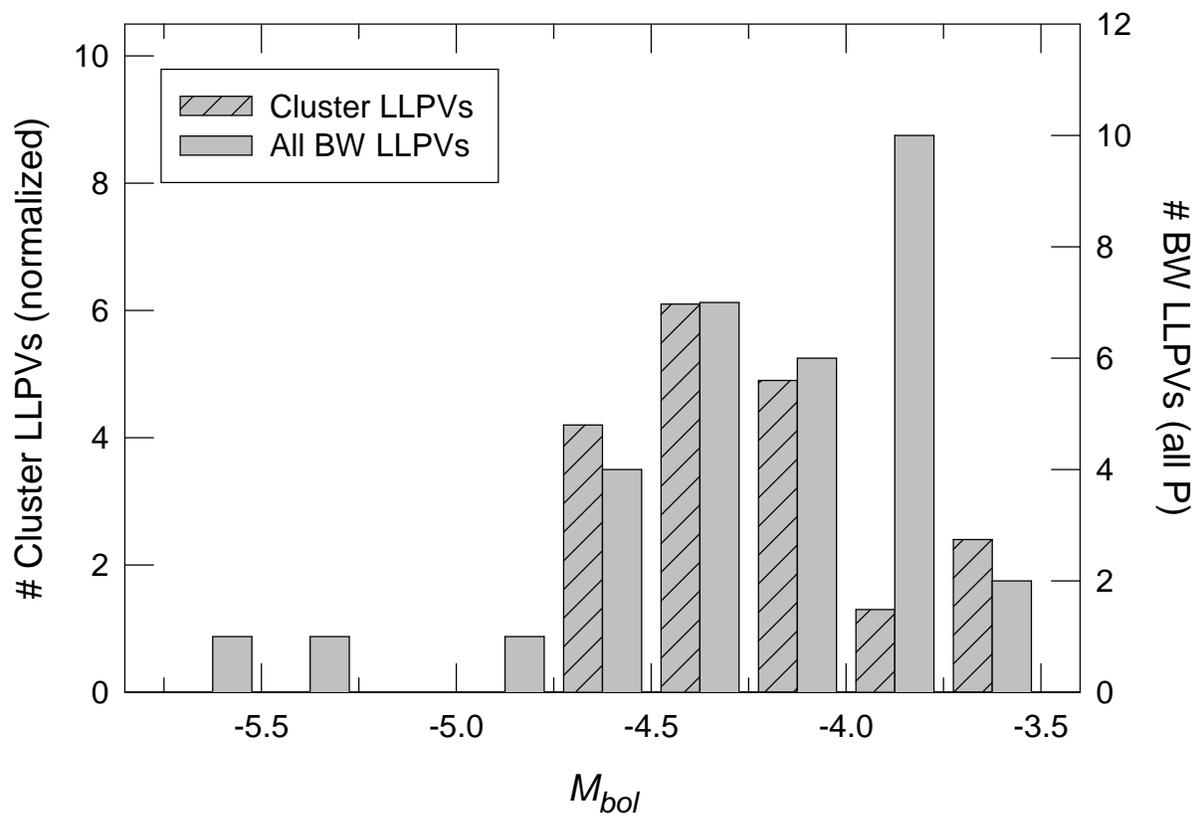

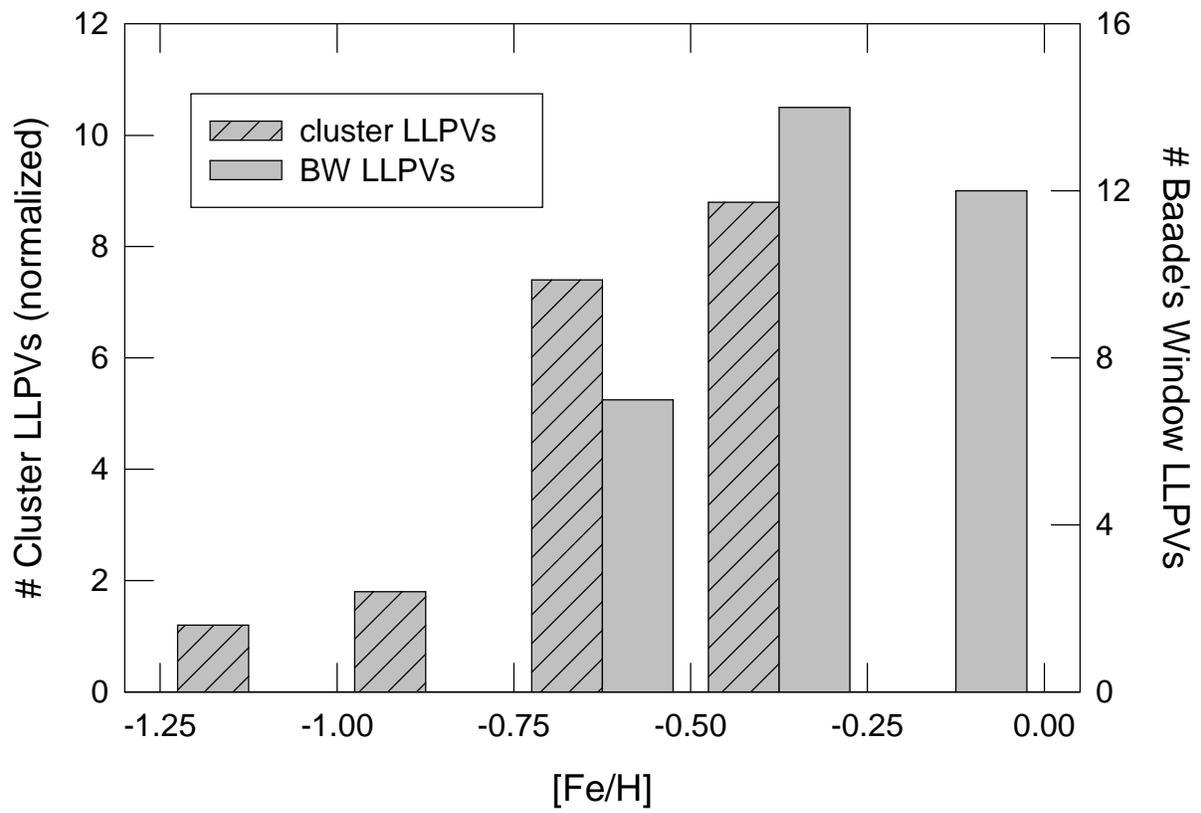

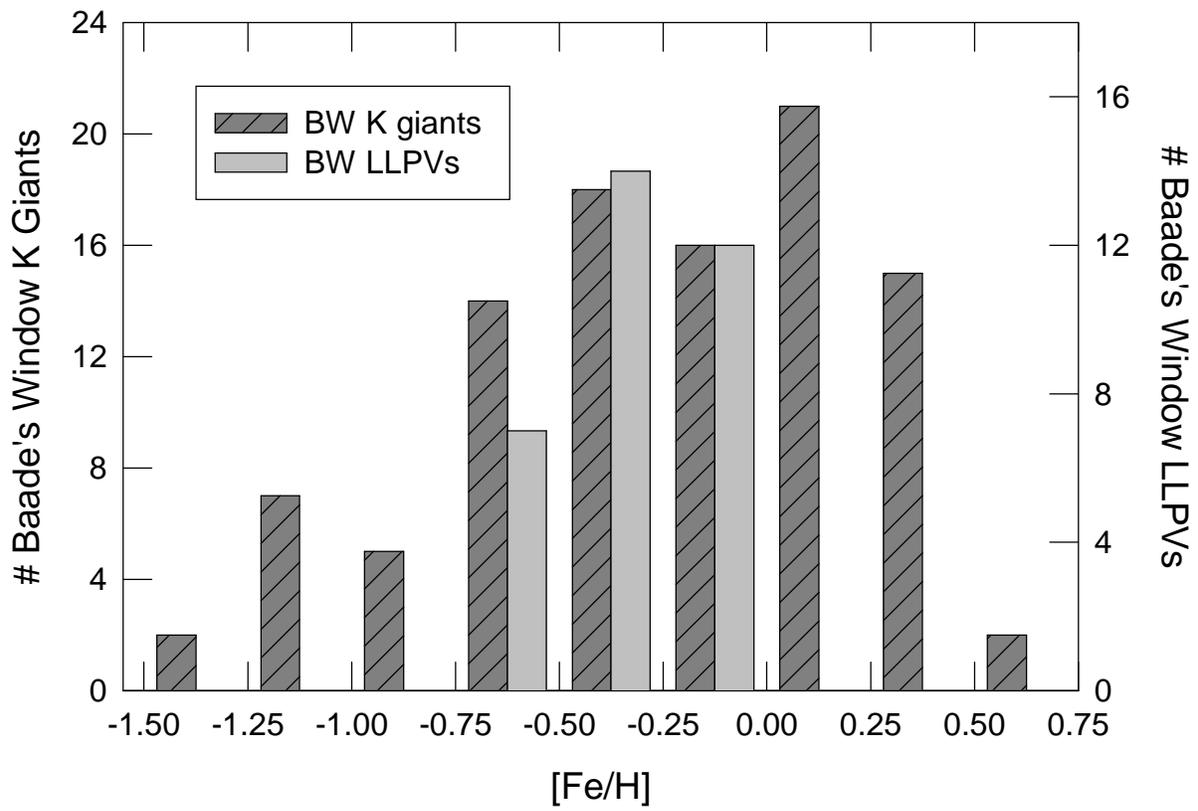

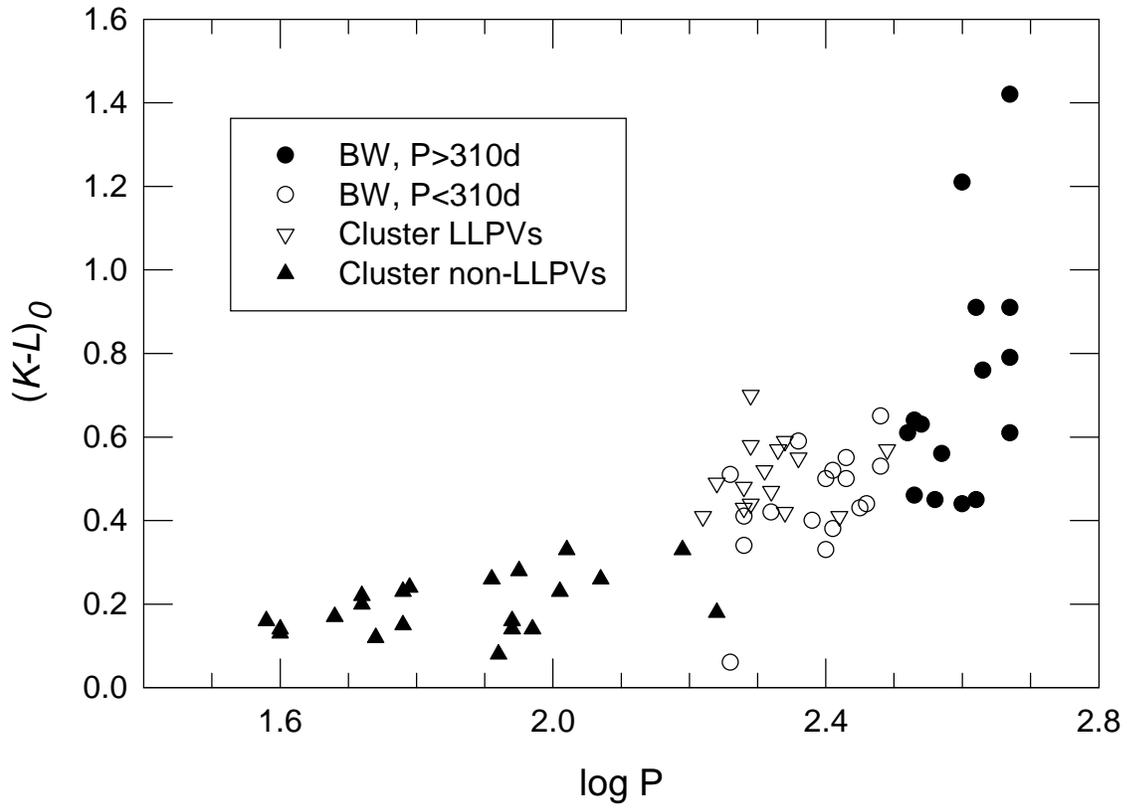

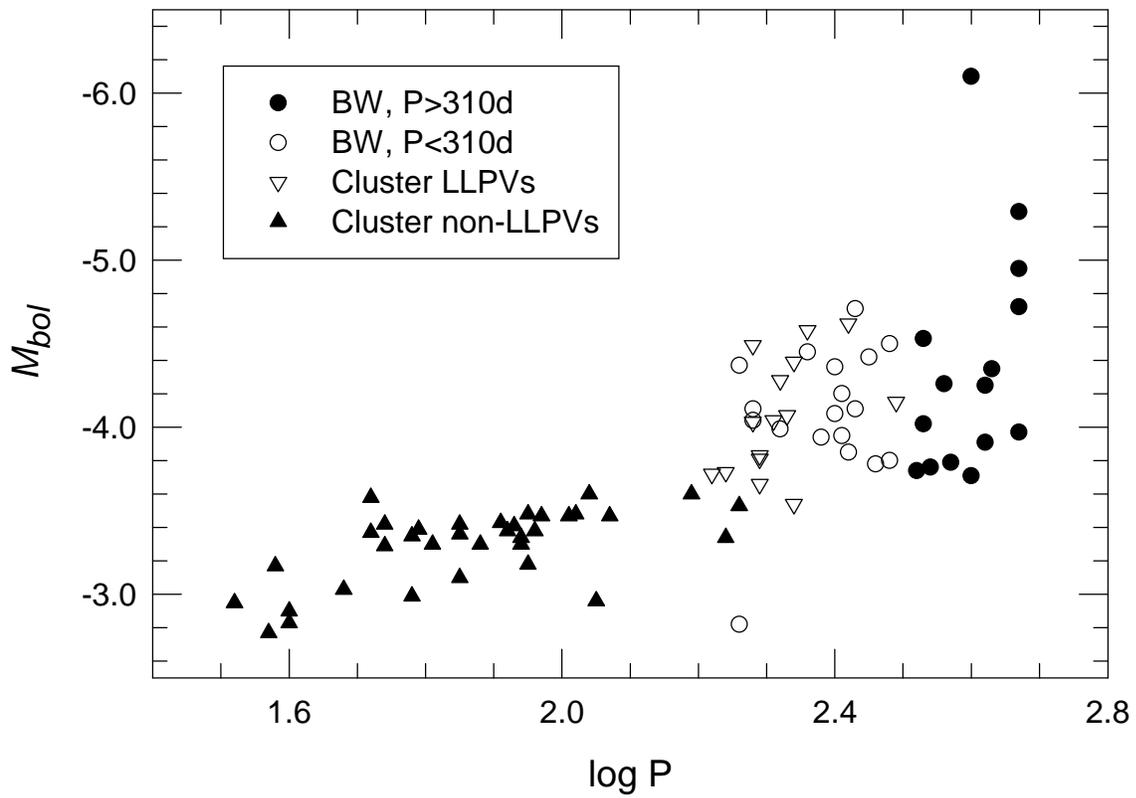